\documentclass[%
 reprint,
 amsmath,amssymb,
 aps,
pra,
]{revtex4-1}

\usepackage{graphicx}% Include figure files
\usepackage{dcolumn}% Align table columns on decimal point
\usepackage{bm}% bold math
\def\ph2{{\it p}-H$_2$}
\begin{document}

%\preprint{APS/123-QED}

\title{Search for superfluidity in supercooled liquid parahydrogen}% Force line breaks with \\

\author{Massimo Boninsegni}
\affiliation{%
Department of Physics, University of Alberta, Edmonton, Alberta, T6G 2E1, Canada}%

\date{\today}% It is always \today, today,

\begin{abstract}
The possible superfluid transition of supercooled liquid parahydrogen is investigated by quantum Monte Carlo simulations. The cooling protocol adopted here allows for the
%investigation
study 
% replaced "investigation" with study
of a fluid phase down to a temperature $T$=0.25 K. No evidence of superfluidity is found, as exchanges of identical particles are  strongly suppressed even at the lowest temperature.  It is shown that, contrary to a commonly held belief, it is not the well depth of the pair-wise interaction but rather its relatively large hard core diameter that physically hinders superfluidity in parahydrogen.

\end{abstract}

\maketitle

%\tableofcontents

\section{\label{intro}Introduction}
A fluid of parahydrogen (\ph2) molecules has long been regarded as a candidate to display a superfluid transition at low temperature, owing to the light mass and bosonic character of its elementary constituents, which are molecules of spin $S=0$. A rough order-of-magnitude estimate of the superfluid transition temperature $T_{\rm c}$ of fluid \ph2, first proposed forty-five years ago \cite{gs}, is based on a very simple model of the system as a gas of non-interacting particles undergoing Bose-Einstein Condensation at $T\sim 6$ K. 
\\ \indent
Despite its crudeness, this approach yields a relatively accurate estimate of $T_{\rm c}$ for the most abundant isotope ($^4$He) of the only known naturally occurring superfluid, namely helium; as first pointed out by Feynman \cite{feynman}, this is because the liquid phase of $^4$He retains the main quality of a non-interacting Bose gas, namely it undergoes BEC at low temperature. The effect of atomic interactions is that of reducing $T_{\rm c}$ from the estimated $3.14$ K to the experimentally observed 2.18 K, at saturated vapor pressure \cite{barenghi}.  
\\ \indent
The experimental observation of the putative superfluid phase of \ph2 is complicated by the fact that, unlike $^4$He, \ph2 freezes into a crystal at a temperature $T$=13.8 K. Although effects of quantum (Bose) statistics are observable in the momentum distribution of liquid \ph2 near melting \cite{boninsegni09}, such a freezing temperature is significantly higher than that at which the speculated BEC would take place in the liquid, according to the above simple argument \cite{note}. 
\\ \indent
On the one hand, attempts to cool a (metastable)  \ph2 fluid down to less than $\sim 8$ K \cite{azuah} have so far not met with success; on the other, it was also suggested by Apenko \cite{apenko} that in the case of \ph2 interactions among molecules may cause a quantitatively more sizable reduction of $T_{\rm c}$ with respect to the free Bose gas estimate, compared to $^4$He, all the way down to $\sim 1$ K. 
More recent theoretical results strongly suggest that even that is likely a significant overestimation. For example,  first principle computer simulations based on realistic intermolecular potentials yield evidence of superfluid behavior at low temperature ($T=1$ K) of nanoscale size  clusters of \ph2 comprising $N \lesssim 20$ molecules \cite{sindzingre,fabio,fabio2,fabio3}, a claim for which some experimental confirmation has been obtained \cite{grebenev,vilesov,li}. Moreover, the strong tendency of the system to form a crystal at low temperature, even in confinement \cite{omiyinka,delmaestro} or in reduced dimensions \cite{boninsegni04,boninsegni13}, greatly reduces the region of parameter space wherein even a metastable fluid phase may exist. 
\\ \indent
What drives \ph2 to crystallize? In a Lennard-Jones type Bose system such as \ph2 or $^4$He,
the interplay of quantum-mechanical and classical effects (the former inhibiting, the latter promoting crystallization)  is embodied in the parameter \cite{deboer}
\begin{equation}
\Lambda = \frac{\hbar^2}{m\epsilon\sigma^2}
\end{equation}
where $m$ is the mass of the constituent particles.
In the $\Lambda\to 0$ limit, the behavior of the system is mostly classical, i.e., quantum-mechanical effects are quantitatively small and/or only observable at very low temperature. Using accepted values \cite{noteacc} for $\epsilon$ and $\sigma$ yields $\Lambda=0.18$ (0.08) for $^4$He (\ph2). Because the values of $\sigma$ are relatively close,  it seems reasonable, and has been customary, to ascribe the crystallization of \ph2 at low temperature {\em mainly} to the depth of the attractive well of the interaction between a pair of \ph2 molecules, which is three times greater than that for two $^4$He atoms. 
It has been recently pointed out, however, that in a Bose system (including $^4$He), the stability against crystallization of a superfluid phase 
crucially hinges on quantum-mechanical exchanges \cite{pollet}; thus, the theoretical question of the relative role of $\epsilon$ and $\sigma$ in suppressing exchanges (hence  promoting crystallization) in \ph2 seems worth revisiting.
\\ \indent
Semiclassical numerical simulations have yielded indirect evidence of a distinct physical behavior of the supercooled fluid, interpreted by the authors as  possible precursor of superfluidity \cite{ando}.
However, the fully-quantum mechanical computer simulation of such a phase by means of,  e.g., quantum Monte Carlo (QMC) methods,  is hampered by the spontaneous crystallization of \ph2 at low temperature \cite{turnbull}, observed at $\lesssim 10$ K even for systems comprising as few as $\sim 100$ molecules and at the freezing density, which is approximately 12\% lower than that of the crystal \cite{notex,superglass}.  For this reason, any attempt to simulate a supercooled phase of \ph2 at sufficiently low temperature by means of equilibrium thermodynamic techniques must resort to {\em ad hoc} tricks, in order to avert the formation of a crystal, at least for a sufficiently long (simulation) time so that meaningful statistics can be collected \cite{superglass}.
\\ \indent
A relatively recent QMC study \cite{soliti}, yielded results interpreted by the authors as evidence  of a possible  metastable superfluid phase of \ph2 at a temperature $\sim$ 1 K, i.e., in ballpark agreement with the prediction of Apenko. The specific simulation procedure adopted therein essentially consists of first weakening and shortening the range of the intermolecular pair potential, which has the effect of stabilizing a superfluid phase by enabling quantum-mechanical exchanges of molecules, and then restoring the potential to the accepted form for \ph2.
The claim made in Ref. \onlinecite{soliti}, is that after restoring the potential a simulated system of relatively small size remains  superfluid, a fact that might be indicative of a long-lived, low-temperature metastable superfluid phase of \ph2. In particular, at a temperature $T$=1 K the reported superfluid fraction $\rho_S$ is as large as $\sim 0.36$, while the Bose condensate fraction $n_\circ$ is $\sim 3.0\times 10^{-4}$, i.e., some two orders of magnitude lower than in $^4$He, for a similar value of $\rho_S$.
\\ \indent
Such a surprising conclusion, difficult to reconcile with theoretical evidence collected in many studies of \ph2 clusters as well as confined \ph2 fluid, seems to warrant further investigation. The most important aspect that needs to be ascertained is the robustness of the results obtained in Ref. \onlinecite{soliti} {\em vis-a-vis} the specific simulation protocol and sampling methodology adopted, as well as the small size of the simulated system, which seems especially important given the very low reported value of $n_\circ$.
\\ \indent 
In order to provide an independent check of the predictions of Ref. \onlinecite{soliti}, we have carried out in this work a QMC study of a metastable fluid phase of \ph2 by means of a different procedure, already utilized in a previous numerical study \cite{superglass} of a possible metastable, overpressurized superfluid phase of $^4$He. 
%Namely, we first equilibrate liquid \ph2 at a temperature $T$=20 K, i.e., above freezing, and then ``quench'' it by restarting the simulation with a lower value of the temperature, in the range between $T$=0.25 K and 1 K. This procedure   allows us to study systems of considerably larger size than those studied in Ref. \onlinecite{soliti}, as the simulated (metastable) liquid phase shows no sign of incipient crystallization over a computer time interval long enough to address satisfactorily at least the most important physical issues.
%\\ \indent 
The results of our study do not support the conclusions reached in Ref. \onlinecite{soliti}. We see no evidence whatsoever of a possible superfluid transition all the way down to $T$=0.25 K, with strong indication that, if a superfluid transition takes place at all, the transition temperature is likely to be at least an order of magnitude lower. The most significant physical feature of the low temperature fluid phase, is the almost complete absence of permutations of molecules, known to underlie BEC and superfluidity. 
%consistently, the one-body density matrix does not flatten off at long distances as reported in Ref. \onlinecite{soliti}, but rather  displays the expected decay, which we observe up to values of the order $10^{-7}$.
\\ \indent
By varying both the well depth and the characteristic diameter of the hard repulsive core of the pair potential at short distance, it is found that it is the latter, not the former, that prevents quantum exchanges from occurring, in the neighborhood of parameter space where \ph2 is situated. Indeed, simulations at $T$=1 K based on the same cooling protocol described above, of a fictitious Bose system with the same particle mass as \ph2 but with a pair potential featuring a well depth equal to one third of the \ph2 interaction also fail to yield any evidence of superfluidity. On the other hand, a  reduction of the value of the hard core diameter by merely $\sim 6\%$, leaving everything else unchanged, immediately causes exchanges to appear, and with those a robust superfluid signal.
\\ \indent 
We have also independently repeated the computer experiment described in Ref. \onlinecite{soliti}, using the same protocol, interaction potentials and number of particles, but, contrary to what claimed therein, we did not find it to be an effective way of stabilizing a metastable superfluid phase of \ph2. For, the artificially created superfluid phase quickly turns crystalline upon restoring the interaction to that between two \ph2 molecules; specifically, long exchange cycles formed in the preliminary stage (i.e., with the altered potential), disappear as quantum-mechanical paths fairly rapidly ``disentangle''. 
%This is observed  regardless of, e.g., how many particles the system comprises, or the shape of the simulation cell. As already reported in previous works, in the absence of the required conditions for \ph2 molecules to arrange themselves in the preferred crystalline structure, the strong tendency to form a crystal results in the formation of whatever structure is allowed by the confining geometry \cite{turnbull}.
The evidence of a possible metastable superfluid phase reported in Ref. \onlinecite{soliti} is therefore quite likely the result of the failure of their sampling procedure to remove all of the exchange cycles created with a weaker interaction, after the pair potential is returned to the form appropriate for \ph2. 
\\ \indent 
The remainder of this article is organized as follows: in Sec. \ref{model} we describe the model and the methodology utilized in this work, we present our results in Sec. \ref{res} and outline our conclusions in Sec. \ref{concl}.
\section{Model and methodology}\label{model}
The system is described as an ensemble of $N$ pointlike, identical particles with  mass equal to that of a \ph2 molecule, and with spin zero, thus  obeying Bose statistics. The system is enclosed in a cubic cell, with periodic boundary conditions in the three directions.  
The quantum-mechanical many-body Hamiltonian reads as follows:
\begin{eqnarray}\label{u}
\hat H = - \lambda \sum_{i}\nabla^2_{i}+\sum_{i<j}v(r_{ij})
\end{eqnarray}
where the first (second) sum runs over all particles (pairs of particles), $\lambda\equiv\hbar^2/2m=12.031$ K\AA$^{2}$, $r_{ij}\equiv |{\bf r}_i-{\bf r}_j|$ and $v(r)$ is a pair potential which describes the interaction between two molecules. We make use in this study of two different forms for $v$, namely the Silvera-Goldman (SG) \cite{SG}, which has been adopted in the vast majority of simulation studies of the condensed phase of \ph2 (including Ref. \onlinecite{soliti}), as well as the simpler Lennard-Jones (LJ) potential, which is more convenient for illustrative purposes \cite{omi}. It need be stated upfront that, at the physical conditions at which simulations are carried out in this work,  these two potentials are essentially equivalent; in particular, {\em all} of the physical conclusions and results presented here are {\em independent} of the specific choice of potential.
\\ \indent 
To the aim of exploring the possible existence of a long-lived (super)fluid phase, at temperatures below the crystallization temperature, we performed first principles QMC simulations of  the system described by Eq.  (\ref{u}), based on the continuous-space Worm Algorithm (WA) \cite{worm,worm2}.  Since this technique is by now fairly well-established, and extensively described in the literature, we shall not review it here. A canonical variant of the algorithm was utilized, in which the total number of particles $N$ is held fixed \cite{fabio,fabio2}.
Details of the simulation are  standard; for instance,  the short-time approximation to the imaginary-time propagator used here is accurate to fourth order in the time step $\tau$ (see, for instance, Ref. \onlinecite{jltp}).  
\\ \indent 
Because the idea is that of utilizing an {\em equilibrium} methodology to investigate the properties of a {\em metastable} phase, it is necessary to adopt an appropriate supercooling protocol, enabling one to collect meaningful statistics for a phase which is not the true equilibrium thermodynamic phase. Our strategy is identical with that of Ref. \onlinecite {superglass}, namely we perform a simulation of the equilibrium  fluid phase of \ph2 at a temperature $T$=20 K and then ``quench'' the system, i.e., restart the simulation from an equilibrated high-temperature configuration with the temperature reset to a lower, ``target'' value. We considered three different target temperatures, namely 0.25, 0.5 and 1 K. 
In order to restart the simulation seamlessly, we leave all the physical parameters unchanged \cite{notey};  in particular, the number of imaginary time slices  is 1280 in {\em all} simulations, including that at high temperature. Thus, the one carried out at the lowest temperature makes use of a time step  $\tau_{\rm \circ} = 3.125\times 10^{-3}$ K$^{-1}$, which has been shown in previous work to render the time step error negligible \cite{boninsegni04}; all other simulations feature a smaller time step than necessary, which does not cause any problem other than rendering the simulations somewhat slower, but still quite manageable on ordinary computing facilities. The simulations described above comprise a number of molecules $N$=512.
The other simulations for which results are offered in sec. \ref {res} for illustrative purposes, based on the SG or LJ potential, were carried out on systems of smaller size (between 128 and 137 molecules), and with a time step equal to $\tau_{\rm \circ}$.

\section{Results}\label{res}
\subsection{Supercooled fluid parahydrogen}

The first test of the effectiveness of the supercooling protocol described in sec. \ref{model}, is whether  indeed it allows us to study the physical properties of a low temperature metastable fluid phase. In order to assess that, we compare first structural properties of the simulated fluid with those of a crystal at the same density, which is set to $\rho=0.023$ \AA$^{-3}$, i.e., the freezing density \cite{you}  of \ph2 at $T$=13.8 K (all of the simulations described in this work are carried out at this density). Although the melting density of \ph2 (0.0261 \AA$^{-3}$) is considerably higher, nonetheless a simulated crystal  at this density (i.e., at negative pressure) is stable against the breakdown of the system into a crystal at the melting density and a low density vapor \cite{notez}. 
\begin{figure}[t]
\centering
\includegraphics[width=\linewidth]{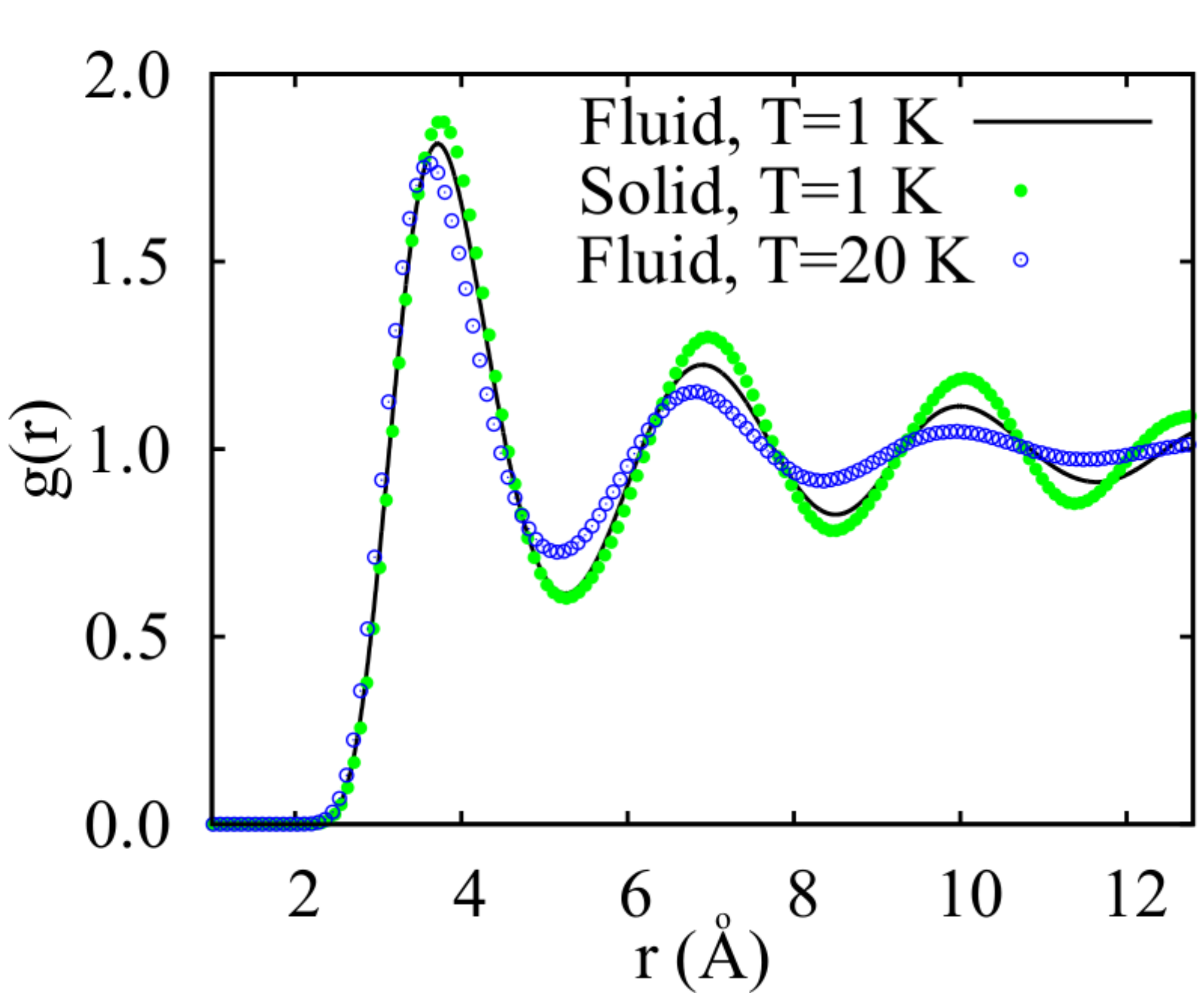}
\caption{{\em Color online.} Pair correlation function $g(r)$ computed by QMC simulation for metastable fluid \ph2 (solid line) and for a {\em hcp} crystal (filled symbols) at $T$=1 K, as well as for the equilibrium fluid at $T$=20 K (open symbols). The density is 0.023 \AA$^{-3}$ and the systems comprises 512 molecules.}
\label{f1}
\end{figure}

Fig. \ref{f1} shows the pair correlation function $g(r)$ for the two phases, at a temperature $T$=1 K; also shown is the same function for the fluid at $T$=20 K, i.e., the system which is then quenched. 
Although the fluid phases displays slightly less pronounced peaks, the quantitative similarity of the three curves at short distance is noteworthy; in particular, the impressive height of the main peak, remarkably  close in all three cases, provides a first indication that the physics of this system is largely dominated by the hard core repulsion between molecules at short distances, to a much more significant degree than, e.g., in solid and liquid helium at comparable thermodynamic conditions (i.e., melting and freezing). 
\\ \indent
It is important to note that the $g(r)$ for the system simulated using the quenching procedure described above (solid curve in Fig. \ref{f1}) is sandwiched between the other two curves, and that its oscillations follow that of the solid only at short distance (i.e., the first three coordination shells around each molecule), much like for the high temperature fluid phase.  
This fact, with the observation that, unlike in other simulations that we describe below, visual inspection of the configurations generated by the sampling algorithm does not show any sign of incipient ordering, constitutes numerical evidence that the cooling protocol adopted in this work indeed affords the investigation of a low temperature, metastable fluid phase of \ph2, i.e., one not featuring any density long range order \cite{notew}.
The physical issue of interest is, of course whether such a metastable phase displays superfluid behavior. 
\\ \indent
The simulation shows that exchanges are {\em exceedingly} infrequent in the metastable fluid phase, essentially as rare as they are in the crystalline one. Crucially, and to the best of our determination, this is not the result of inadequate sampling. Indeed, the updates that within the WA  lead to the appearance of long exchange cycles (``swap'' updates \cite{worm2}) are accepted at a rate (roughly around 25\% over a portion of world line of length 0.05 K$^{-1}$, see for details Ref. \onlinecite{worm2}) comparable to that of a superfluid $^4$He simulation. Yet, it is extremely rare for single-particle world lines not to close onto themselves. We interpret this as a physical result, genuinely reflecting the underlying character of the simulated system, rather than a numerical artifact.
\begin{figure}[t]
\centering
\includegraphics[width=\linewidth]{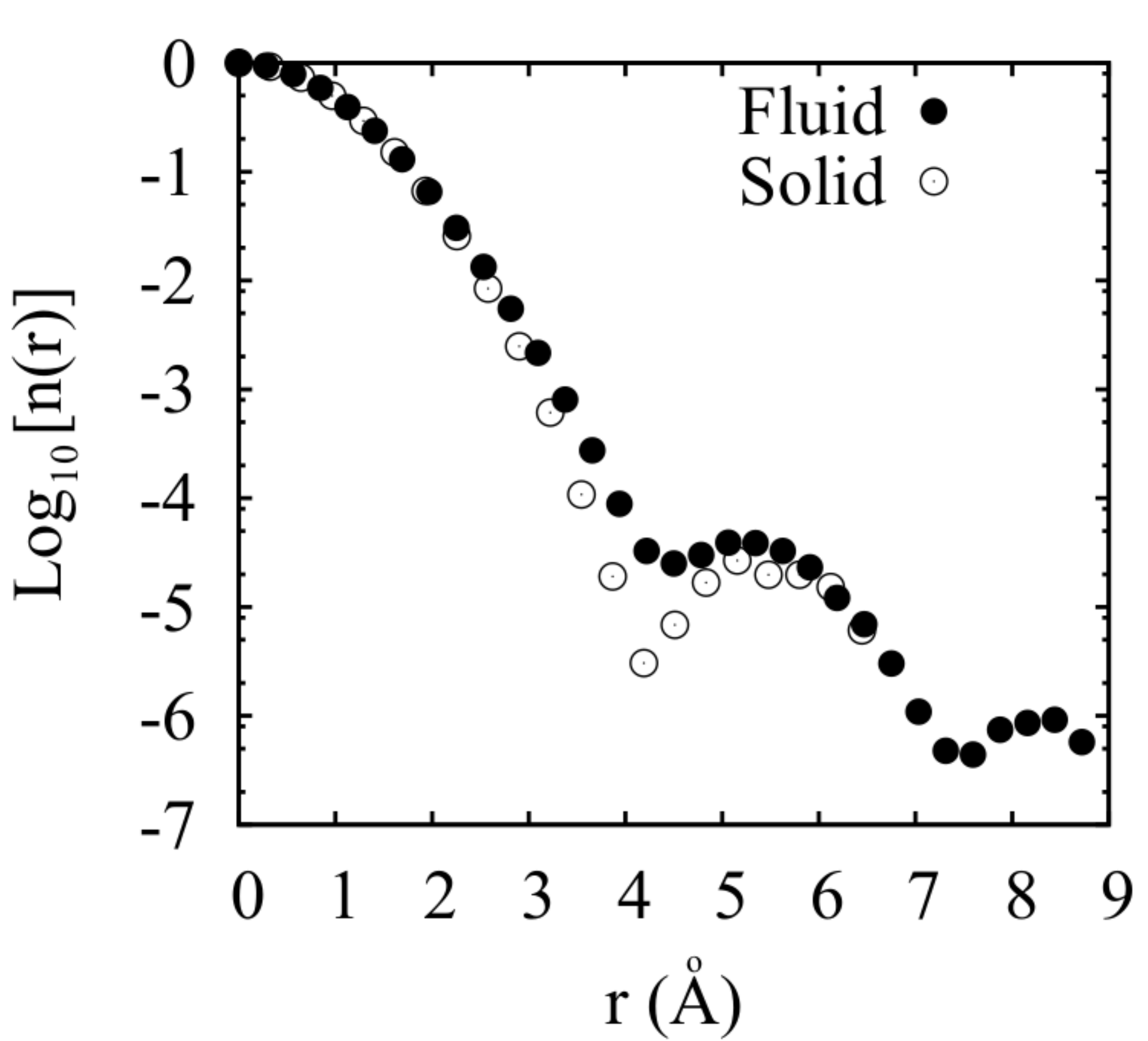}
\caption{ One-body density matrix $n(r)$ computed by QMC simulation for metastable fluid \ph2 (solid circles) and for a {\em hcp} crystal (open circles), both at density 0.023 \AA$^{-3}$. The temperature is $T$=1 K. Statistical errors are, at  most, the size of the symbols.}
\label{f2}
\end{figure} 
\\ \indent
Consequently to the lack of exchanges, the system fails to develop off-diagonal long range order and superfluidity, as shown by the exponential decay at long distance of the one-particle density matrix (Fig. \ref{f2}). The comparison with the same quantity computed for the solid at the same density only shows an enhancement in the fluid at distances of the order of the mean interparticle separation; otherwise the two curves are virtually indistinguishable at short distances and beyond the first coordination shell, within the statistical errors of the calculation.
\\ \indent
No significant quantitative changes are seen on lowering the temperature to $T$=0.5 and $T$=0.25 K. Indeed, exchanges remain strongly suppressed even at the lowest temperature, and no detectable difference can be seen in the computed one-body density matrix with respect to that shown in Fig. \ref{f2}, within statistical errors.
\\ \indent Thus, the main physical conclusion of this numerical investigation is that if a metastable fluid phase of \ph2 can be stabilized at low temperature, it will not turn superfluid down to a temperature $T$=0.25 K; moreover, the lack of any appreciable temperature dependence of the results, especially of the propensity of the system to quantum-mechanical exchanges, suggests that no metastable superfluid phase may be observable at all, as in the low temperature limit only the crystalline phase is expected,
as in lower dimensions \cite{boninsegni04,boninsegni13}. This is clearly very different from what observed in the case of $^4$He; here, the absence of exchanges and the concomitant failure of the system to develop density long-range order suggest that the physics of this metastable \ph2 fluid to be closer to that of a conventional (i.e., non-superfluid) glass.
\\ \indent 
This conclusion is of course at variance with that of Ref. \onlinecite{soliti}, in which simulations of a possible low temperature metastable fluid phase were carried out using a very different cooling protocol; we come back to this point below. Next, we discuss the physical reasons as to why this system, long deemed a plausible candidate for superfluidity, fails instead to display even a hint of it, both experimentally as well as in first principle calculations.

\subsection{What hinders superfluidity?}
In order to explore in detail the intriguing issue of absence of superfluidity in the simulated metastable fluid phase, we performed additional simulations, using the same cooling protocol described above, of systems featuring the mass of \ph2 molecules but with modified intermolecular pair potentials.  
This analysis is more conveniently carried out using the LJ potential, owing to its simpler form. Parameters of the various simulations are summarized in Table \ref{tableone}.
\\ \indent
We started out with a fictitious system (henceforth referred to as A) featuring the \ph2 hard core diameter, namely $\sigma=2.96$ \AA, but with a well depth $\epsilon=10.22$ K, i.e., that of $^4$He, which is roughly a factor three smaller than that of \ph2, with a corresponding threefold increase of the value of $\Lambda$, to approximately 0.27, i.e., considerably greater than that of $^4$He. Unlike \ph2, system A has a superfluid  ground state, whose equilibrium density can be estimated \cite{zillich} at less than 0.01 \AA$^{-3}$. Its thermodynamic equilibrium phase is however a crystal, if compressed to the density of this study, namely 0.023 \AA$^{-3}$. \\ \indent
\begin{table}
\begin{ruledtabular}
\begin{tabular}{cccccc}
System &$\epsilon$ (K) & $\sigma$ (\AA) &$\Lambda$ &$\rho_\circ $ (\AA$^{-3}$) &SF \\ \hline
A &10.22 & 2.96 &0.27 & $\lesssim 0.010$ &no\\
B &34.16 & 2.75 &0.09 &$\sim 0.025$ & yes\\
$^4$He &10.22 & 2.556&0.18 &0.022 &yes  \\
\end{tabular}
\end{ruledtabular}
\caption{\label{table1} Parameters for the different Lennard-Jones systems simulated. Systems A and B comprise particles with mass equal to that of \ph2 molecules. Also shown are the values of the de Boer parameter $\Lambda$, approximate values for the $T=0$ equilibrium density $\rho_\circ$, as well as whether the system is superfluid at $T=1$ K at the density $\rho=0.023$ \AA$^{-3}$. Last row reports for comparison the values of the parameters for $^4$He, the value of $\Lambda$ reflecting the greater mass of the $^4$He atoms. }\label{tableone}
\end{table}
When simulating a metastable fluid phase of A, we found no evidence of superfluidity down to $T$=0.25 K, even though A displays a more pronounced fluidlike behavior than \ph2 (the height of the first peak of the $g(r)$ is close to 1.6). In particular, no significant increase is observed in the frequency with which exchanges occur.
Thus, the superfluid transition temperature of this metastable fluid phase is in this case too considerably lower than $T$=0.25 K. The same conclusion was reached by simulating a system interacting via a modified version of the SG potential, with a well depth three times smaller than the actual one. 
\\ \indent 
We then simulated a second system (B), in which the well depth was left at its regular value (i.e., $\epsilon=34.16$ K) but the hard core diameter reduced to $\sigma=2.75$ \AA, i.e., by approximately 6\%; as a result, the $\Lambda$ parameter only increases to $\sim$ 0.09 (from 0.08 for \ph2). Clearly, this fictitious system is physically much closer to \ph2, in parameter space, than system A. Its ground state is a crystal, with an equilibrium density above 0.023 \AA$^{-3}$, but below that of \ph2, namely 0.0261 \AA$^{-3}$. Simulations carried out at $T$=1 K yield a strong superfluid signal (the superfluid fraction is 100\% within statistical errors); in fact, in this case no special cooling protocol is needed, as the system immediately develop long exchanges, even if the simulation is started out from a lattice arrangement.
\begin{figure}[t]
\centering
\includegraphics[width=\linewidth]{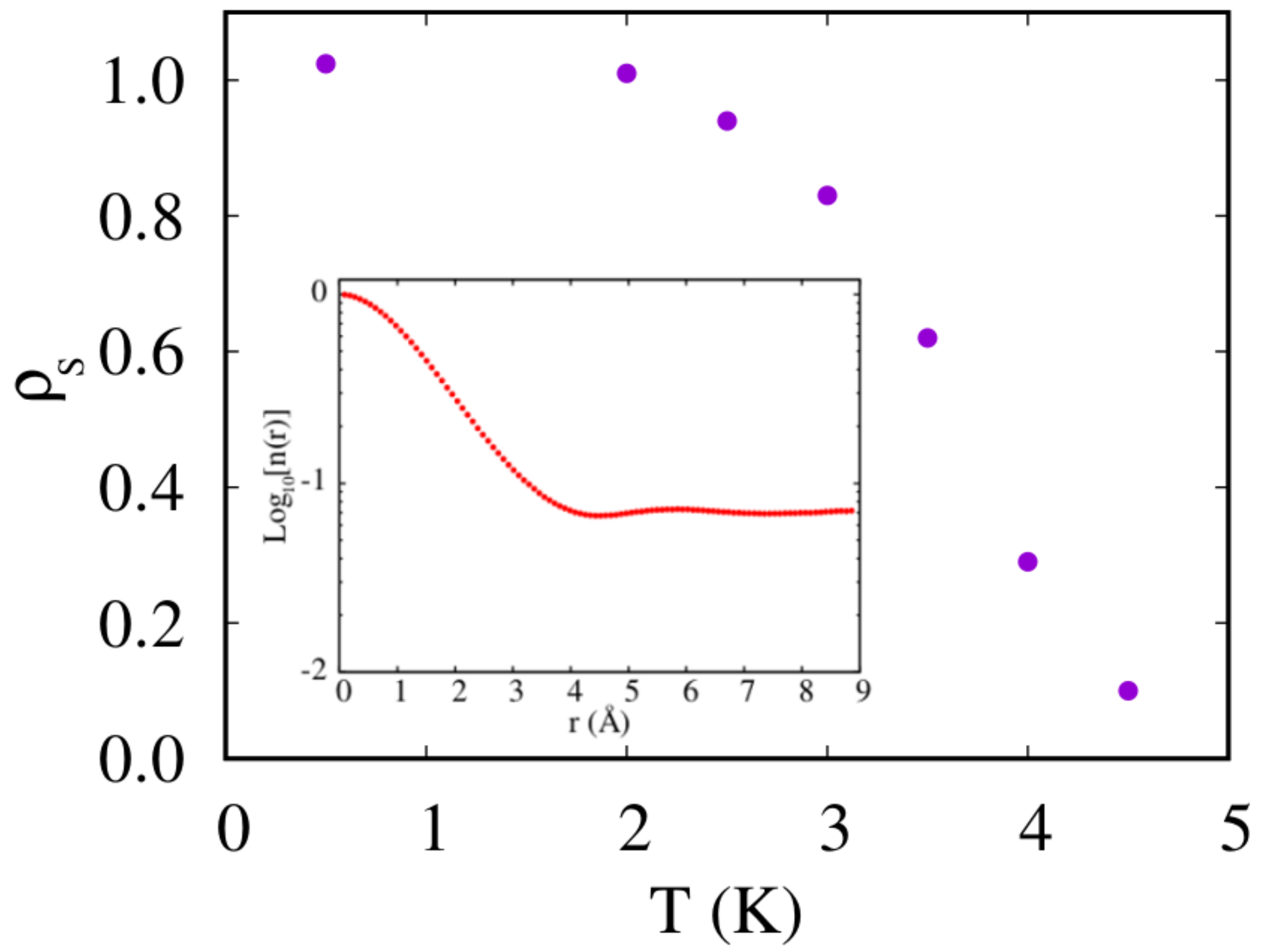}
\caption{{\em Color online}. Superfluid fraction $\rho_S(T)$ versus temperature for an ensemble of 128 particles with the mass of \ph2 molecules at a density $\rho=0.023$ \AA$^{-3}$, interacting  via a LJ potential with well depth $\epsilon=34.16$ K and hard core diameter $\sigma=2.556$ \AA. Inset displays the one-body density matrix $n(r)$ at $T$=1 K. Statistical errors are at the most of the size of the symbols.}
\label{f3}
\end{figure} 
\indent
If $\sigma$ is reduced even further, to 2.556 \AA\ (i.e., the $^4$He value,  a 13\% reduction compared to the \ph2 one), as a result of which $\Lambda\sim 0.11$, still closer to the \ph2 than to the $^4$He value, the superfluid transition occurs at a temperature $\gtrsim 4$ K \cite{noteext}, i.e., considerably above that of $^4$He, as shown in Fig. \ref{f3}. In this case, the one-body density matrix flattens off at long distances as expected, saturating to a value close to 7\% in the low temperature limit.
\\ \indent All of this constitutes strong numerical evidence that \ph2, at least in parameter space, is not as far from being a superfluid as may have been thought. Indeed, a relatively small increase in the value of the parameter $\Lambda$ would lead to a superfluid at least as robust as $^4$He. What is perhaps surprising is the observation  that, while the well depth of the interaction potential between two \ph2 molecules has traditionally been believed to be the main obstacle to the appearance of a superfluid phase, the results of this simulation suggest instead that, at the typical thermodynamic conditions of liquid \ph2 (near freezing) it is mostly the relatively large diameter of the repulsive core at short distance that has the most significant effect in preventing exchanges of molecules, thereby preventing superfluidity and also strengthening the crystalline phase \cite{pollet}.
\begin{figure}[t]
\centering
\includegraphics[width=\linewidth]{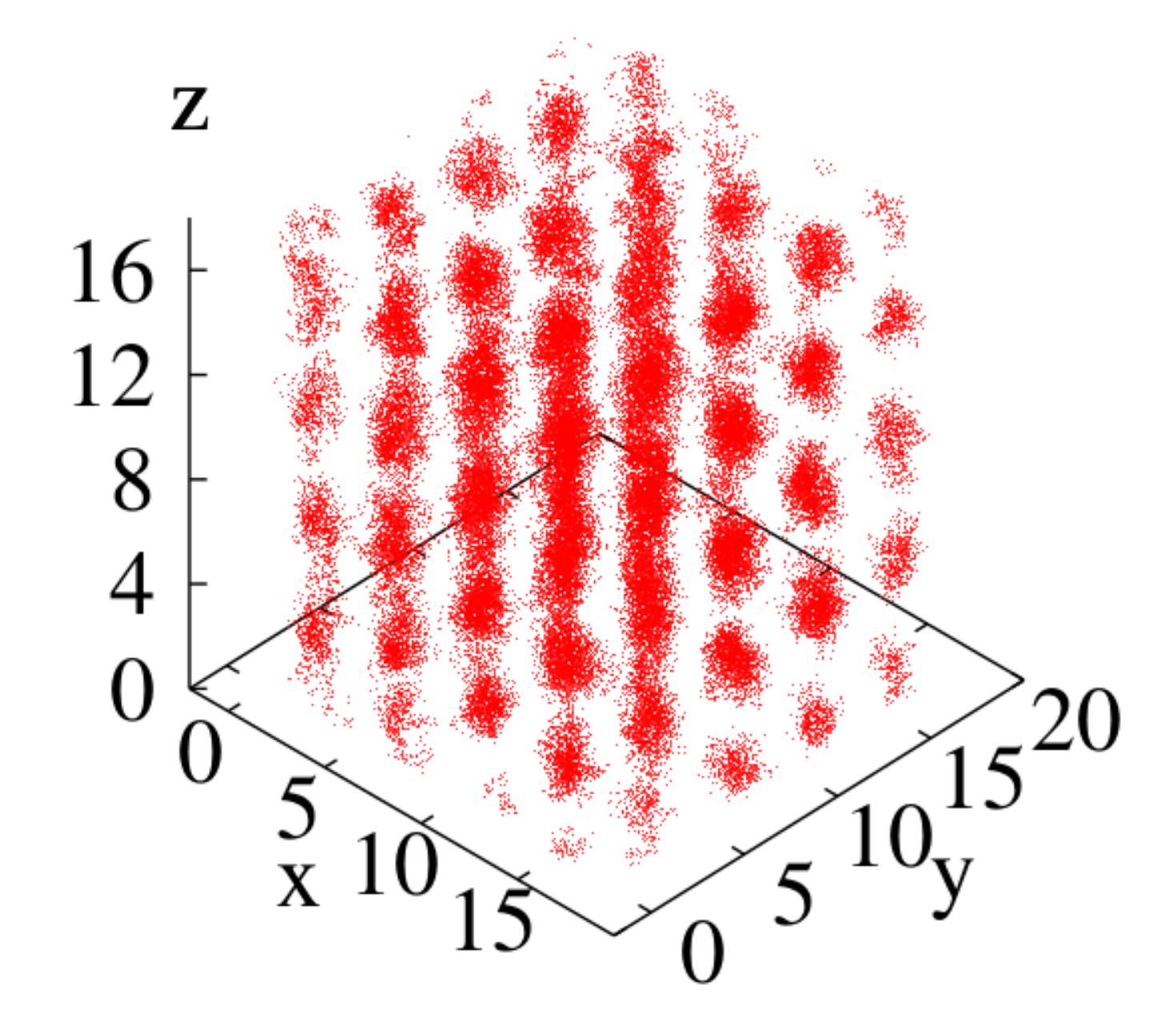}
\caption{{\em Color online.} Configurational snapshot (particle worldlines, lengths are in \AA) of a system of $N$=130 \ph2 molecules interacting via the SG pair potential. The system is enclosed in a cubic box. The temperature is $T$=1 K and the density $\rho$=0.0234 \AA$^{-3}$. The system was initially equilibrated to a superfluid phase using the Aziz pair potential, as suggested in Ref. \onlinecite{soliti}.}
\label{f4}
\end{figure} 
\\ \indent 
We now discuss the disagreement between the results and conclusions of this work and those of Ref. \onlinecite{soliti}. We have attempted to duplicate their numerical experiment by first stabilizing a superfluid phase using a different potential, either the $^4$He Aziz potential or a LJ potential with $\epsilon=34.16$ K and $\sigma=2.556$ \AA, which as we have shown above also leads to superfluidity, and then restoring the actual \ph2 pair interaction (either SG or LJ).
As mentioned above, we did not find this protocol to be effective in stabilizing anything that could be regarded as meaningfully representative of a metastable superfluid phase. In actuality, as long as the crucial ``swap'' moves are attempted sufficiently often, 
the long exchange cycles established with the modified pair potential disappear fairly quickly, with a corresponding decrease to zero of the initial superfluid signal. As soon as the system rids itself of long permutation cycles, it quickly begins to develop solid order, in qualitative  agreement with the findings of Ref. \onlinecite{pollet}. This happens {\em regardless of how many particles and/or shape of the simulation cell}. 
%On this point, it is worth noting that it is a common, naive misconception that one may be able to circumvent crystallization merely by using a cell shape or number of particles incompatible with the preferred crystalline structure. In those cases, as noticed time and again \cite{turnbull,boninsegni16}, the system simply does the ``next best thing'', namely form whatever crystal (possibly with defects) is allowed by the geometry of the simulation.
An example is shown in Fig. \ref{f4}, displaying an instantaneous many-particle configuration arising from a simulation performed as suggested in Ref. \onlinecite{soliti}, namely using a cubic box and $N$=130 \ph2 molecules, equilibrating first a superfluid phase at $T$=1 K using the $^4$He Aziz pair potential and then switching back to the SG. The system immediately starts crystallizing, not necessarily in any of the three main directions of the cell (obviously), but nonetheless in a way clearly identifiable by straightforward visual inspection of the configurations generated  by the sampling.
 \\ \indent
As mentioned above, as the system reverts to its equilibrium, crystalline phase, the superfluid signal decreases to zero. The same outcome is observed in simulations comprising fewer particles, e.g., with $N=90$, a number used in Ref. \onlinecite{soliti}.
It is possible, and we have observed this in our simulations too, to have at times few isolated and resilient permutation cycles, which if the system is sufficiently small can wind around the periodic boundaries, giving rise to a finite superfluid signal. However, such a signal is spurious, not meaningfully representative of any observable physical phenomenon but rather of underlying sampling issues. It is easy to detect this anomaly by histogramming the frequency of occurrence $P(n)$ of exchange cycles including $1\le n\le N$ molecules; this is a smoothly decaying function in a superfluid, whereas the presence of few isolated peaks at specific numbers generally signals  the inability of the underlying sampling procedure to remove all cycles. It seems quite likely that this may be the reason for the alleged evidence of superfluid behavior reported in Ref. \onlinecite{soliti}. 
\section{Conclusions} \label{concl}
We have carried out extensive, first principle numerical simulations, aimed at assessing the possibility that a long-lived metastable fluid phase of \ph2 may exist, down to a temperature of the order of 1 K, where it is expected to undergo a superfluid transition. By using a specific cooling protocol, we were able to stabilize a phase displaying fluidlike properties all the way down to a temperature $T$=0.25 K. We have observed a strong suppression of exchanges of identical particles in this system, even at this low temperature, and the consequent absence of any hint of superfluid behavior.
\\ \indent On performing simulations with slightly varied intermolecular pair potentials, we obtained results suggesting that liquid \ph2 at the freezing density is not actually far, parameter-wise, from displaying the originally predicted superfluid behavior; in particular, it is found that, at its specific thermodynamic conditions, superfluidity in \ph2 is primarily inhibited by the relatively large range of the repulsive core of the pairwise potential, more than by the depth of the attractive well, to which the inability of \ph2 to display a fluid phase at low temperature has been traditionally attributed. In general, both potential strength and radius, in conjunction with particle mass, intrinsically affect the quantum-mechanical behavior of a many-body system, not just for Lennard-Jones type interactions \cite{saccani}. Because of the subtle but crucial role played by exchanges of identical particles, not only in enabling superfluidity but also in opposing crystallization, criteria to assess the proximity of a system to a superfluid phase based on a single parameter (like $\Lambda$ in this case) are not dependable.
\\ \indent 
We have attempted to repeat the simulation work of Ref. \onlinecite {soliti}, suggesting that a metastable superfluid phase of \ph2 may exist at $T$=1 K. 
The results presented here invalidate such a contention, which is likely a result of sampling problems and simulations on systems of very small size.
\begin{acknowledgments}
This work was supported in part by the Natural Sciences and Engineering Research Council of Canada (NSERC). The author gratefully acknowledges the hospitality of the International Centre for Theoretical Physics, Trieste, where most of this research work was carried out. Computing support of Westgrid is also acknowledged.
\end{acknowledgments}


\begin{thebibliography}{99}
\bibitem{noteacc}
Specifically, for $^4$He it is $\epsilon=10.22$ K and $\sigma=2.556$ \AA; for \ph2, $\epsilon=34.16$ K and $\sigma=2.96$ \AA.
\bibitem{notex}
This behavior is in stark contrast with the case of $^4$He. See, for instance, Ref. \onlinecite{superglass}.
\bibitem{notew}
Obviously, given a sufficient simulation time crystal ordering begins to emerge. Indeed, we have found 512 particles to be the minimum system size required in order for our simulation protocol to be effective, i.e., for the disordered phase to be sufficiently resilient. In simulations carried out on smaller systems, crystal ordering arises before meaningful averages of physical estimates can be collected.
\bibitem{noteext}
A careful assessment of the transition temperature obviously requires a rigorous finite size scaling analysis. It was not pursued in this work because it is not necessary to establish the main conclusion.
\end{thebibliography}
\end{document}